\documentclass{ws-ijmpa}

\begin{document}

\markboth{Dariusz Prorok} {Multiplicities in p-p($\bar{p}$)
collisions and NBD}

%
\catchline{}{}{}{}{}
%

\title{MULTIPLICITIES IN
ULTRARELATIVISTIC PROTON-(ANTI)PROTON COLLISIONS AND NEGATIVE
BINOMIAL DISTRIBUTION FITS
%
}

\author{DARIUSZ PROROK
%
}

\address{Institute for Theoretical Physics, University of
Wroc{\l}aw\\ Pl.Maksa Borna 9, 50-204 Wroc{\l}aw, Poland
\\prorok@ift.uni.wroc.pl}

\maketitle

\begin{history}
\received{Day Month Year}
\revised{Day Month Year}
\end{history}

\begin{abstract}
Likelihood ratio tests are performed for the hypothesis that
charged-particle multiplicities measured in proton-(anti)proton
collisions at $\sqrt{s}$ = 0.9 and 2.36 TeV are distributed
according to the negative binomial form. Results indicate that the
hypothesis should be rejected in the all cases of ALICE-LHC
measurements in the limited pseudo-rapidity windows, whereas should
be accepted in the corresponding cases of UA5 data. Possible
explanations of that and of the disagreement with the least-squares
fitting method are given.

\end{abstract}

\ccode{PACS numbers: 13.85.Hd, 29.85.Fj}

\section{Introduction}
\label{intro}

The UA5 Collaboration noticed for the first time that
charged-particle multiplicity distributions measured in high energy
proton-(anti)proton collisions in limited intervals of
pseudo-rapidity have the negative binomial form \cite{Alner:1985zc}.
In the present paper this observation will be verified for the
collisions at $\sqrt{s}$ = 0.9 and 2.36 TeV performed by UA5
\cite{Ansorge:1988kn} and ALICE Collaborations \cite{Aamodt:2010ft}.
Only non-single diffractive (NSD) events will be considered because
such a case was analyzed with this respect by both Collaborations.
In fact, the author investigated ALICE inelastic events also
(including the case of $\sqrt{s}$ = 7 TeV \cite{Aamodt:2010pp}), but
all fits were entirely unacceptable.

The Negative Binomial Distribution (NBD) is defined as

\begin{equation}
P(n; p, k) = \frac{k(k+1)(k+2)...(k+n-1)}{n!} (1-p)^{n}p^{k} \;,
\label{NBDist}
\end{equation}

\noindent where $n = 0, 1, 2,...$, $0 \leq p \leq 1$ and $k$ is a
positive real number. In the application to high energy physics $n$
has the meaning of the number of charged particles detected in an
event. The expected value $\bar{n}$ and variance $V(n)$
\footnote{Here, these quantities are distinguished from the
experimentally measured the average charged particle multiplicity
$\langle N_{ch} \rangle$ and the variance $\sigma^{2}$.
\label{przyp1}} are expressed as:

\begin{equation}
\bar{n} = \frac{k(1-p)}{p}\;,\;\;\;\;\;\;\;\;V(n) =
\frac{k(1-p)}{p^{2}} \;. \label{Parametpk}
\end{equation}

In this analysis the hypothesis that the charged-particle
multiplicities measured in high energy $p-p(\bar{p})$ collisions are
distributed according to the NBD is verified with the use of the
maximum likelihood method (ML) and the likelihood ratio test. More
details of this approach can be found in
Refs.~\refcite{Cowan:1998ji}--\refcite{Baker:1983tu}.

There are two crucial reasons for this approach:

\begin{romanlist}[(ii)]
 \item The fitted quantity is a probability distribution function (p.d.f.),
 so the most natural way is to use the ML method, where the
 likelihood function is constructed directly from the tested p.d.f..
 But more important is that because of Wilks's theorem (see Appendix
 B) one can easily define a statistic, the distribution of which converges to a $\chi^2$
 distribution as the number of measurements goes to infinity. Thus
 for the large sample the goodness-of-fit can be expressed as a
 $p$-value computed with the corresponding $\chi^2$ distribution.

 \item The most commonly used method, the least-squares method (LS)
 (called also $\chi^2$ minimization), has the disadvantage of
 providing only the qualitative measure of the significance of the
 fit, in general. Only if observables are represented by Gaussian
 random variables with known variances, the conclusion about
 the goodness-of-fit equivalent to that mentioned in the point (i)
 can be derived \cite{Cowan:1998ji}.

\end{romanlist}

It is worth noting that the ML method with binned data and Poisson
fluctuations within a bin was already applied to fitting
multiplicity distributions to the NBD but at much lower energies
(E-802 Collaboration \cite{Abbott:1995as}).

\section{The maximum likelihood method}
\label{Likmeth}

The number of charged particles $N_{ch}$ is assumed to be a random
variable with the p.d.f. given by Eq.~(\ref{NBDist}). Each event is
treated as an independent observation of $N_{ch}$ and a set of a
given class of events is a sample. For $N$ events in the class there
are $N$ measurements of $N_{ch}$, say $\textbf{X} = \{
X_1,X_2,...,X_N \}$. Some of these measurements can be equal, {\it
i.e.} $X_i = X_j$ for $i \neq j$ can happen. The whole population
consists of all possible events with the measurements of 0, 1, 2,...
charged particles and by definition is infinite \footnote{
Precisely, because of the energy conservation the number of produced
charged particles is limited but the number of collisions is not.
\label{przyp2}}.

For the class of events one can defined the likelihood function

\begin{equation}
L(\textbf{X} \mid p, k) = \prod_{j=1}^{N}\; P(X_j; p, k) \;,
\label{Likelfun}
\end{equation}
%
where $P(X_j; p, k)$ is the NBD, Eq.~(\ref{NBDist}).

The values $\hat{p}$ and $\hat{k}$ for which $L(\textbf{X} \mid p,
k)$ has its maximum are the maximum likelihood estimates of
parameters $p$ and $k$. This is equivalent to the maximization of
the log-likelihood function

\begin{equation}
\ln L(\textbf{X} \mid p, k) = \sum_{j=1}^{N}\;\ln P(X_j; p, k) \;.
\label{logLikfun}
\end{equation}

Thus the values $\hat{p}$ and $\hat{k}$ are the solutions of the
equations:

\[
\frac{\partial}{\partial p}\ln L(\textbf{X} \mid p, k) =
\sum_{j=1}^{N}\;\frac{\partial}{\partial p} \ln P(X_j; p, k)=0 \;,
\label{logLikeqp}
\]
\begin{equation}
\frac{\partial}{\partial k}\ln L(\textbf{X} \mid p, k) =
\sum_{j=1}^{N}\;\frac{\partial}{\partial k} \ln P(X_j; p, k)=0 \;.
\label{logLikeqk}
\end{equation}

It can be proven that one of the necessary conditions for the
existence of the maximum is (see Appendix A for details):

\begin{equation}
\bar{n} = \langle N_{ch} \rangle \;, \label{eqalaver}
\end{equation}
{\it i.e.} the distribution average has to be equal to the
experimental average.

\section{Likelihood ratio test}
\label{Liktest}

Let divide the sample defined in Sect.~\ref{Likmeth} into $m$ bins
characterized by $Y_i$ - the number of measured charged particles
\footnote{Now $Y_i \neq Y_j$ for $i \neq j$ and $i, j = 1, 2,...,m$.
\label{przyp3}} and $n_i$ - the number of entries in the $i$th bin,
$N = \sum_{i=1}^{m}\;n_i$ (details of the theoretical framework of
this Section can be found in
Refs.~\refcite{Cowan:1998ji}--\refcite{Baker:1983tu}). Then the
expectation value of the number of events in the $i$th bin can be
written as

\begin{equation}
\nu_i(\nu_{tot},p,k) = \nu_{tot} \cdot P(Y_i; p, k) \;,
\label{Neventi}
\end{equation}
where $\nu_{tot}$ is the expected number of all events in the
sample, $\nu_{tot} = \sum_{i=1}^{m}\;\nu_i$. This is because one can
treat the number of events in the sample $N$ also as a random
variable with its own distribution - Poisson one. Generally, the
whole histogram can be treated as one measurement of $m$-dimensional
random vector $\textbf{n}=(n_1,...,n_m)$ which has a multinomial
distribution, so the joint p.d.f. for the measurement of $N$ and
$\textbf{n}$ can be converted to the form
\cite{Cowan:1998ji,Baker:1983tu}:

\begin{equation}
f(\textbf{n};\nu_1,...,\nu_m) = \prod_{i=1}^{m}
\frac{\nu_i^{n_i}}{n_i!}\;\exp{(-\nu_i)} \;. \label{jointpdf}
\end{equation}
Since now $f(\textbf{n};\nu_1,...,\nu_m)$ is the p.d.f. for one
measurement, $f$ is also the likelihood function

\begin{equation}
L(\textbf{n} \mid \nu_1,...,\nu_m) = f(\textbf{n};\nu_1,...,\nu_m)
\;. \label{Ljointpdf}
\end{equation}
With the use of Eq.~(\ref{Neventi}) the corresponding likelihood
function can be written as

\begin{equation}
L(\textbf{n} \mid \nu_{tot}, p, k) = L(\textbf{n} \mid
\nu_1(\nu_{tot},p,k),...,\nu_m(\nu_{tot},p,k)) \;. \label{Liksubset}
\end{equation}
Then the likelihood ratio is defined as

\begin{equation}
\lambda = \frac{L(\textbf{n} \mid \hat{\nu}_{tot}, \hat{p},
\hat{k})}{L(\textbf{n} \mid \breve{\nu}_1,...,\breve{\nu}_m)} =
\frac{L(\textbf{n} \mid \hat{\nu}_{tot}, \hat{p},
\hat{k})}{L(\textbf{n} \mid n_1,...,n_m)} \;. \label{Likeliratio}
\end{equation}
where $\hat{\nu}_{tot}$, $\hat{p}$ and $\hat{k}$ are the ML
estimates of $\nu_{tot}$, $p$ and $k$ with the likelihood function
given by Eq.~(\ref{Liksubset}) and $\breve{\nu}_i = n_i$, $i = 1,
2,...m$ are the ML estimates of $\nu_i$ treated as free parameters.
Note that since the denominator in Eq.~(\ref{Likeliratio}) does not
depend on parameters, the log-ratio defined as

\begin{eqnarray}
&&\ln{\lambda(\nu_{tot}, p, k)} = \ln { \frac{L(\textbf{n} \mid
\nu_{tot}, p, k)}{L(\textbf{n} \mid n_1,...,n_m)} } \cr \cr & & = -
\sum_{i=1}^{m}\;\bigg (n_i \ln{\frac{n_i}{\nu_i}} + \nu_i - n_i
\bigg ) \cr \cr & & = - \nu_{tot} + N - \sum_{i=1}^{m}\;n_i
\ln{\frac{n_i}{\nu_i}} \;, \label{Logratio}
\end{eqnarray}
where $\nu_i$ are expressed by Eq.~(\ref{Neventi}), can be used to
find the ML estimates of $\nu_{tot}$, $p$ and $k$. Further, the
statistic given by

\begin{equation}
\chi^2 = -2 \ln \lambda = 2 \sum_{i=1}^{m}\;\bigg (n_i
\ln{\frac{n_i}{\hat{\nu}_i}} + \hat{\nu}_i - n_i \bigg ) \;
\label{PoissonChi}
\end{equation}
approaches the $\chi^2$ distribution asymptotically, {\it i.e.} as
the number of measurements, here the number of events $N$, goes to
infinity (the consequence of the Wilks's theorem, see Appendix B).
The values $\hat{\nu}_i$ are the estimates of $\nu_i$ given by

\begin{equation}
\hat{\nu}_i = \hat{\nu}_{tot} \cdot P(Y_i; \hat{p}, \hat{k}) \;
\label{Nevenhati}
\end{equation}
and if one assumes that $\nu_{tot}$ does not depend on $p$ and $k$
then $\hat{\nu}_{tot} = N$. For such a case

\begin{equation}
\sum_{i=1}^{m} \hat{\nu}_i = \sum_{i=1}^{m} n_i \; \label{Nevenexp}
\end{equation}
and Eq.~(\ref{PoissonChi}) becomes

\begin{equation}
\chi^2(\hat{p}, \hat{k}) = -2 \ln \lambda = 2 \sum_{i=1}^{m}\; n_i
\ln{\frac{n_i}{\hat{\nu}_i}}.  \; \label{MultinomChi}
\end{equation}
%
Also then one can just put $\nu_{tot} = N$ and Eq.~(\ref{Logratio})
can be rewritten as

\begin{eqnarray}
&&\ln{\lambda(p, k)} \cr \cr & & = N \cdot \ln{N} -
\sum_{i=1}^{m}\;n_i \ln{n_i} + \sum_{i=1}^{m}\;n_i \ln{P(Y_i; p, k)}
\cr \cr & & = - \sum_{i=1}^{m}\;n_i \ln{\frac{n_i}{N}} + N
\sum_{i=1}^{m}\;\frac{n_i}{N} \ln{P(Y_i; p, k)} \cr \cr & & = - N
\sum_{i=1}^{m}\;P_i^{ex} \ln{P_i^{ex}} + N \sum_{i=1}^{m}\;P_i^{ex}
\ln{P(Y_i; p, k)}, \label{Logratfreq}
\end{eqnarray}
with the term depending on $p$ and $k$ the same as
Eq.~(\ref{LogLikfubi}) and $P_i^{ex} = n_i/N$. Therefore conclusions
of Appendix A holds here, {\it i.e.} the necessary conditions for
the existence of the maximum is $\bar{n} = \langle N_{ch} \rangle$,
Eq.~(\ref{DlogLikfk}) is the equation which determines $\hat{k}$ and
$\hat{p}$ is obtained with the help of Eq.~(\ref{Oneoverp}). Note
that the maximum of $\ln{\lambda}$ is the minimum of $\chi^2 = -2 \;
\ln{\lambda}$, so from Eqs.~(\ref{MultinomChi}) and
(\ref{Logratfreq}) one arrives at

\begin{equation}
\chi_{min}^2 = -2\; N \sum_{i=1}^{m}\;P_i^{ex} \ln{\frac{P(Y_i;
\hat{p}, \hat{k})}{P_i^{ex}}} \;. \label{FinalChi}
\end{equation}
In fact, the method just described assumes that the sum of $P_i(p,
k) \equiv P(Y_i; p, k)$ over all bins equals 1. But only the
infinite sum of $P(n; p, k)$ is 1. However the measured values of
$Y_m$ are big enough (of the order of 20 at least for all considered
cases) so in the vicinity of $\hat{p}$ and $\hat{k}$ the sum of
$P(n; p, k)$ up to $n = Y_m$ equals 1 approximately (see the seventh
column in Table~\ref{Table1}). Nevertheless, to calculate
$\chi_{min}^2$, Eq.~(\ref{FinalChi}), $P(Y_i; \hat{p}, \hat{k})$
were normalized appropriately and these results are listed in the
fifth column of Tables~\ref{Table1}-\ref{Table3}. Another way to
treat this problem is to create arbitrary the $(m+1)$st bin for all
possible $n
> Y_m$ and with $P_{m+1}^{ex} = 0$. Bins with $P_i^{ex} = 0$ ($n_i =
0$ equivalently) do not contribute in Eq.~(\ref{FinalChi}) (see
Ref.~\refcite{Cowan:1998ji}). In practice, it means that
$\chi_{min}^2$ would be calculated also from Eq.~(\ref{FinalChi})
but without the normalization. It has turned out that that way leads
to much greater values of $\chi_{min}^2$.

\section{Results and discussion}
\label{Finl}

The method described in Sections~\ref{Likmeth} and \ref{Liktest}
requires that all bins in a given data set have the width equal to
1, so as the experimental probability $P_i^{ex}$ to measure a signal
in the $i$th bin was equivalent to the probability of the
measurement of $(i-1)$ charged particles (the first bin is the bin
of 0 charged particles detected). This is fulfilled for all bins of
the considered data sets except the ends of their tails. In these
tails the measured values of $P_i^{ex}$ have been uniformly
distributed over the bin range so as the method could be applied
directly. If the bin width is not significantly greater than 1 then
this approximation should not change substantially the value of
$\chi_{min}^2$ given by Eq.~(\ref{FinalChi}) because in the most
cases $P_i^{ex}$ at tails are two orders smaller than in the main
part of distributions. Also errors in tails are bigger, in the range
$10-50 \% $, increasing with $i$.

Since the test statistic $-2 \ln \lambda$ has a $\chi^{2}$
distribution approximately in the large sample limit, it can be used
as a test of the goodness-of-fit. The result of the test is given by
the so-called $p$-value which is the probability of obtaining the
value of the statistic, Eq.~(\ref{PoissonChi}), equal to or greater
then the value just obtained by the ML method for the present data
set, when repeating the whole experiment many times:

\begin{equation}
p = P(\chi^{2} \geq  \chi_{min}^{2}; n_d) =
\int_{\chi_{min}^{2}}^{\infty}\;f(z;n_d) dz \;, \label{Pvalue}
\end{equation}
where $f(z;n_d)$ is the $\chi^{2}$ p.d.f. and $n_d$ the number of
degrees of freedom, $n_d = m-2$ here.

\begin{table}[!]
\tbl{Results of fitting multiplicity distributions for the NSD
events measured in $p-\bar{p}$ (UA5) $^{2}$ and $p-p$ (ALICE) $^{3}$
collisions. The ALICE numbers of events are from Ref.~9. The
distributions have been modified in the tails so as all bins have
the width 1, see the text for explanations.}
{
\begin{tabular}{@{}cccccccccc@{}}\toprule
 & & & & $\chi^2$/$n_d$ & & &
\multicolumn{3}{c}{$\chi_{LS}^2$/$n_d$ with errors:}
\\
 \cline{8-10} Experiment & N & $\hat{k}$ & $\hat{p}$ & $\chi^2$ & p-value &
 $\sum P_i(\hat{p}, \hat{k})$ & quadrature & sum & statistical
\\
$\sqrt{s}$ & & & & ($n_d$)& [\%]& & sum & & only
\\
\colrule
 UA5 & 8550.0 & 1.5574
& 0.3012 & 0.339 & 99.97 & 0.99996 & 0.375 & na & na
\\
$\sqrt{s}= 0.9$ TeV & (80 \% \textit{eff}.) & $\pm 0.0365$ & $\pm
0.0054$ & 10.16 & & & & &
\\
$\mid \eta \mid <$ 0.5 &  & & & (30) & & & & &
\\
 & & & & &
\\
\hline
 UA5 & 10000.0 & 1.5574
& 0.3012 & 0.396 & 99.87 & 0.99996 & 0.375 & na & na
\\
$\sqrt{s}= 0.9$ TeV & (70 \% \textit{eff}.) & $\pm 0.0337$ & $\pm
0.0050$ & 11.88 & & & & &
\\
$\mid \eta \mid <$ 0.5 &  & & & (30) & & & & &
\\
 & & & & &
\\
\hline
 ALICE & 149663.16 &
1.3764 & 0.2767 & 14.155 & 0 & 0.99960 & 1.116 & 0.576 & 3.089
\\
$\sqrt{s}= 0.9$ TeV & & $\pm 0.0076$ & $\pm 0.0012$ & 353.88 & & & &
&
\\
$\mid \eta \mid <$ 0.5 & & & & (25) & & & & &
\\
 & & & & &
\\
 ALICE & 128476.45 &
1.4316 & 0.1625 & 37.761 & 0 & 0.99865 & 1.886 & 1.034 & 11.51
\\
$\sqrt{s}= 0.9$ TeV & & $\pm 0.0070$ & $\pm 0.0008$ & 1548.21 & & &
& &
\\
$\mid \eta \mid <$ 1.0 & & & & (41) & & & & &
\\
 & & & & &
\\
 ALICE & 60142.77 &
1.4955 & 0.1332 & 22.051 & 0 & 0.99876 & 2.993 & 1.671 & 15.31
\\
$\sqrt{s}= 0.9$ TeV & & $\pm 0.0102$ & $\pm 0.0009$ & 1168.69 & & &
& &
\\
$\mid \eta \mid <$ 1.3 & & & & (53) & & & & &
\\
 & & & & &
\\
\hline
 UA5 & 8550.0 & 1.7987 & 0.1385 & 0.812 & 87.81 & 0.99991 & 0.487 &
na & na
\\
$\sqrt{s}= 0.9$ TeV & (80 \% \textit{eff}.) & $\pm 0.0319$ & $\pm
0.0024$ & 60.12 & & & & &
\\
$\mid \eta \mid <$ 1.5 & & & & (74) & & & & &
\\
 & & & & &
\\
\hline
 UA5 & 10000.0 & 1.7987 & 0.1385 & 0.950 & 59.99 & 0.99991 & 0.487 &
na & na
\\
$\sqrt{s}= 0.9$ TeV & (70 \% \textit{eff}.) & $\pm 0.0295$ & $\pm
0.0022$ & 70.31 & & & & &
\\
$\mid \eta \mid <$ 1.5 & & & & (74) & & & & &
\\
 & & & & &
\\
\hline
 ALICE & 38970.79 &
1.1778 & 0.2084 & 6.266 & 0 & 0.99930 & 0.888 & 0.501 & 3.592
\\
$\sqrt{s}= 2.36$ TeV & & $\pm 0.0115$ & $\pm 0.0018$ & 194.26 & & &
& &
\\
$\mid \eta \mid <$ 0.5 & & & & (31) & & & & &
\\
 & & & & &
\\
 ALICE & 37883.99 &
1.2139 & 0.1180 & 17.416 & 0 & 0.99726 & 2.209 & 1.312 & 17.73
\\
$\sqrt{s}= 2.36$ TeV & & $\pm 0.0103$ & $\pm 0.0010$ & 853.37 & & &
& &
\\
$\mid \eta \mid <$ 1.0 & & & & (49) & & & & &
\\
 & & & & &
\\
 ALICE & 22189.40 &
1.2123 & 0.0927 & 15.561 & 0 & 0.99644 & 4.0557 & 2.4537 & 34.40
\\
$\sqrt{s}= 2.36$ TeV & & $\pm 0.0129$ & $\pm 0.0010$ & 949.22 & & &
& &
\\
$\mid \eta \mid <$ 1.3 & & & & (61) & & & & &
\\
 & & & & &
\\
\botrule
\end{tabular} \label{Table1}}
\end{table}

\begin{table}[!]
\tbl{Results of testing the NBD for the original data sets of the
NSD events measured in $p-\bar{p}$ (UA5) $^{2}$ and $p-p$ (ALICE)
$^{3}$ collisions. The ALICE numbers of events are from Ref.~9. The
values of $\hat{k}$ and $\hat{p}$ are taken from Table~\ref{Table1}.
}
{
\begin{tabular}{@{}cccccccccc@{}} \toprule
 & & & & $\chi^2$/$n_d$ & &
\multicolumn{4}{c}{$\chi_{LS}^2$/$n_d$ with errors:}
\\
 \cline{7-10} Experiment & N & $\hat{k}$ & $\hat{p}$ & $\chi^2$ & p-value
 & quadrature & sum & statistical & $\sim\sqrt{n_i}$
\\
$\sqrt{s}$ & & & & ($n_d$)& [\%]& sum & & only &
\\ \colrule
 UA5 & 8550.0 & 1.5574
& 0.3012 & 0.211 & 99.998 & 0.072 & na & na & 0.203
\\
$\sqrt{s}= 0.9$ TeV & (80 \% \textit{eff}.) & $\pm 0.0365$ & $\pm
0.0054$ & 4.859 & & & & &
\\
$\mid \eta \mid <$ 0.5 &  & & & (23) & & & & &
\\
 & & & & &
\\
\hline
 UA5 & 10000.0 & 1.5574
& 0.3012 & 0.247 & 99.991 & 0.072 & na & na & 0.237
\\
$\sqrt{s}= 0.9$ TeV & (70 \% \textit{eff}.) & $\pm 0.0337$ & $\pm
0.0050$ & 5.683 & & & & &
\\
$\mid \eta \mid <$ 0.5 &  & & & (23) & & & & &
\\
 & & & & &
\\
\hline
 ALICE & 149663.16 &
1.3764 & 0.2767 & 14.498 & 0 & 0.728 & 0.381 & 2.458 & 15.107
\\
$\sqrt{s}= 0.9$ TeV & & $\pm 0.0076$ & $\pm 0.0012$ & 347.95 & & & &
&
\\
$\mid \eta \mid <$ 0.5 & & & & (24) & & & & &
\\
 & & & & &
\\
 ALICE & 128476.45 &
1.4316 & 0.1625 & 36.855 & 0 & 1.718 & 0.948 & 11.010 & 38.017
\\
$\sqrt{s}= 0.9$ TeV & & $\pm 0.0070$ & $\pm 0.0008$ & 1547.91 & & &
& &
\\
$\mid \eta \mid <$ 1.0 & & & & (42) & & & & &
\\
 & & & & &
\\
 ALICE & 60142.77 &
1.4955 & 0.1332 & 24.323 & 0 & 2.213 & 1.276 & 15.201 & 25.771
\\
$\sqrt{s}= 0.9$ TeV & & $\pm 0.0102$ & $\pm 0.0009$ & 1167.51 & & &
& &
\\
$\mid \eta \mid <$ 1.3 & & & & (48) & & & & &
\\
 & & & & &
\\
\hline
 UA5 & 8550.0 & 1.7987 & 0.1385 & 1.099 & 28.94 & 0.362 &
na & na & 1.14
\\
$\sqrt{s}= 0.9$ TeV & (80 \% \textit{eff}.) & $\pm 0.0319$ & $\pm
0.0024$ & 57.16 & & & & &
\\
$\mid \eta \mid <$ 1.5 & & & & (52) & & & & &
\\
 & & & & &
\\
\hline
 UA5 & 10000.0 & 1.7987 & 0.1385 & 1.286 & 8.06 & 0.362 &
na & na & 1.33
\\
$\sqrt{s}= 0.9$ TeV & (70 \% \textit{eff}.) & $\pm 0.0295$ & $\pm
0.0022$ & 66.85 & & & & &
\\
$\mid \eta \mid <$ 1.5 & & & & (52) & & & & &
\\
 & & & & &
\\
\hline
 ALICE & 38970.79 &
1.1778 & 0.2084 & 7.030 & 0 & 0.761 & 0.428 & 3.805 & 7.465
\\
$\sqrt{s}= 2.36$ TeV & & $\pm 0.0115$ & $\pm 0.0018$ & 189.82 & & &
& &
\\
$\mid \eta \mid <$ 0.5 & & & & (27) & & & & &
\\
 & & & & &
\\
 ALICE & 37883.99 &
1.2139 & 0.1180 & 18.535 & 0 & 2.288 & 1.362 & 18.802 & 20.282
\\
$\sqrt{s}= 2.36$ TeV & & $\pm 0.0103$ & $\pm 0.0010$ & 852.59 & & &
& &
\\
$\mid \eta \mid <$ 1.0 & & & & (46) & & & & &
\\
 & & & & &
\\
 ALICE & 22189.40 &
1.2123 & 0.0927 & 18.233 & 0 & 4.245 & 2.599 & 39.647 & 19.980
\\
$\sqrt{s}= 2.36$ TeV & & $\pm 0.0129$ & $\pm 0.0010$ & 948.11 & & &
& &
\\
$\mid \eta \mid <$ 1.3 & & & & (52) & & & & &
\\
 & & & & &
\\
\botrule
\end{tabular} \label{Table2}}
\end{table}

The results of the analysis are presented in Table~\ref{Table1}.
Note that for UA5 cases two possibilities of the corrected number of
events are listed. This is because only the measured number of
events, $6839$, is given in Ref.~\refcite{Ansorge:1988kn}. However,
the fits have been done to the corrected distributions, so also the
corrected number of events should be put into Eq.~(\ref{FinalChi}).
The number have been estimated in the following way: in Fig.4 of
Ref.~\refcite{Ansorge:1988kn} the mean of the observed distribution
versus the corrected (true) number of particles is plotted, the
curve is a straight line roughly with the tangent equal to $\sim
0.8$, so one can guess that the efficiency is also about $80 \%$.
Just to check how results are stable with respect to a change in the
number of events, the case with $70 \%$ efficiency has been also
calculated. As one can see, for all ALICE cases the hypothesis in
question should be rejected, whereas for the listed UA5 cases should
be accepted. But it was claimed that charged-particle multiplicities
measured in the limited pseudo-rapidity windows by the ALICE
Collaboration are distributed according to the NBD
\cite{Aamodt:2010ft,Aamodt:2010pp,Mizoguchi:2010vc}. However that
conclusion was the result of the $\chi^2$ minimization (the LS
method). Therefore it seems to be reasonable to check what are the
values of the LS $\chi^2$ function at the ML estimators listed in
the third and fourth columns of Table~\ref{Table1}. For the sample
described in Sect.~\ref{Liktest} one can define the LS $\chi^{2}$
function as:

\begin{equation}
\chi_{LS}^{2}(p,k) = \sum_{i=1}^{m} \frac{(P_i^{ex}-P(Y_i; p, k)
)^{2}}{err_{i}^{2}}   \;, \label{Chidef}
\end{equation}

\noindent where $err_{i}$ is the uncertainty of the $i$th
measurement. Here this function \textbf{is not minimized} with
respect to $p$ and $k$ as in the LS method but is calculated at ML
estimates of $p$ and $k$, {\it i.e.} at $\hat{p}$ and $\hat{k}$. One
can see from the eight and ninth columns of Table~\ref{Table1} that
$\chi_{LS}^2$/$n_d$ values are significant for the ALICE narrowest
pseudo-rapidity windows, what agrees with the results of
Ref.~\refcite{Mizoguchi:2010vc}.

Since the determination of $\hat{k}$ and $\hat{p}$ has been done for
the distributions modified in their tails, as it has been just
explained, one should check what values of $\chi^2$ and
$\chi_{LS}^2$ are at $\hat{k}$ and $\hat{p}$ for the original data
sets. It means that if the $i$th bin width is greater than 1,
instead of $P(Y_i; \hat{p}, \hat{k})$ in Eq.~(\ref{FinalChi}) the
appropriate sum $\sum\;P(n; \hat{p}, \hat{k})$ over $n \in
\textrm{bin}\; i$ is taken. The results of the check are presented
in Table~\ref{Table2}. Qualitatively the results are the same as in
Table~\ref{Table1}, only slight differences in numbers can be
noticed except the UA5 cases (for $\mid \eta \mid <$ 0.5 $\chi^2$
has decreased more than 2 times, but the change is in the good
direction). This is because the maximal width of a tail bin is 2 for
all ALICE cases, but is 8 and 17 for UA5 windows $\mid \eta \mid <$
0.5 and $\mid \eta \mid <$ 1.5, respectively. Of course, the
assumption of the uniform distribution inside a wider bin causes
greater discrepancies. Nevertheless, the results of the test for
both UA5 cases are positive even if ($\hat{k}$, $\hat{p}$) is not
the maximum of the exact likelihood function (in fact, values of
$\hat{k}$ are the same as those obtained by UA5 Collaboration in
Ref.~\refcite{Ansorge:1988kn}). This is guaranteed by the Wilks's
theorem (see Appendix B), which allows for the test of a single
point in the parameter space. Then the tested point might not be the
best estimate of the true value but the hypothesis in question
becomes the hypothesis only about a particular distribution (a
\textit{simple} hypothesis). This is also the reason why $n_d = m$
in Table~\ref{Table2}. In terms of rigorous statistics single points
are tested in there.

In all ALICE cases $\chi^2$ values listed in the fifth column of
Table~\ref{Table2} are only slightly smaller than corresponding ones
from Table~\ref{Table1}. For $\mid \eta \mid <$ 0.5 the decrease is
about $2 \%$, for other cases is less than $0.1 \%$. Also
$\chi^2$/$n_d$ values are much greater than 1. Therefore it is
reasonable to recognize $\hat{k}$ and $\hat{p}$ determined for
modified data sets as a good approximations of the ML estimators.
Thus the hypothesis about the NBD should be rejected on the basis of
obtained values of $\chi^2$/$n_d$ and $p$-values.

One can also compare $\chi^2$/$n_d$ with $\chi_{LS}^2$/$n_d$
calculated for the original data sets and the same $\hat{k}$ and
$\hat{p}$. The results are listed in four last columns of
Table~\ref{Table2} for various treatment of errors. Note that for
UA5 conclusions from both statistics are exactly the same. In the
ALICE both cases of the window $\mid \eta \mid <$ 0.5,
$\chi_{LS}^2$/$n_d < 1$ is acceptable for errors expressed as the
quadrature sum of statistical and systematical components and is
smaller than the corresponding values in Table~\ref{Table1}. In
other ALICE cases $\chi_{LS}^2$/$n_d$ is substantially greater than
1 for the same treatment of errors. This is in the full agreement
with the results of Ref.~\refcite{Mizoguchi:2010vc}. One can also
check what $\chi_{LS}^2$/$n_d$ is if only statistical errors are
taken into account. The results are listed in the next to last
column of Table~\ref{Table2}. For all ALICE cases the values are
much greater than 1. This means that acceptable $\chi_{LS}^2$/$n_d$
was obtained only because of significant systematic errors of ALICE
measurements. The word ''significant'' is subjective, here means
''significant with respect to the sample size'', not to the value of
$P_i^{ex}$.

The crucial question is now why the conclusions from $\chi^2$ and
$\chi_{LS}^2$ test statistics are the same for UA5 data but entirely
opposite for ALICE measurements? The main difference between both
statistics is that $\chi^2$ depends explicitly on the number of
events but $\chi_{LS}^2$ does not. On opposite, $\chi^2$ does not
depend on the actual errors but $\chi_{LS}^2$ does. In fact,
$\chi^2$ statistic implicitly assumes errors of the type
$\sqrt{n_i}$, what is the straightforward result of the form of the
likelihood function, Eqs.~(\ref{jointpdf}) and (\ref{Ljointpdf}),
namely the product of Poisson distributions. This is revealed when
one compare $\chi^2$/$n_d$ and $\chi_{LS}^2$/$n_d$ with errors $\sim
\sqrt{n_i}$ (the fifth and last column in Table~\ref{Table2}). The
values are practically the same.

\begin{table}[!]
\tbl{Results of testing the NBD for the original data sets of the
NSD events measured in $p-\bar{p}$ (UA5) $^{2}$ and $p-p$ (ALICE)
$^{3}$ collisions at $\sqrt{s}= 0.9$ TeV. The values of $\hat{k}$
and $\hat{p}$ are taken from Table~\ref{Table1} but the ALICE
numbers of events have been changed arbitrarily to the UA5 number of
events. }
{
\begin{tabular}{@{}cccccccccc@{}} \toprule & & & & $\chi^2$/$n_d$ & &
\multicolumn{4}{c}{$\chi_{LS}^2$/$n_d$ with errors:}
\\
 \cline{7-10} Experiment & N & $\hat{k}$ & $\hat{p}$ & $\chi^2$ & p-value
 & quadrature & sum & statistical & $\sim\sqrt{n_i}$
\\
$\sqrt{s}$ & & & & ($n_d$)& [\%]& sum & & only &
\\
 \colrule
 UA5 & 8550.0 & 1.5574
& 0.3012 & 0.211 & 99.998 & 0.072 & na & na & 0.203
\\
$\sqrt{s}= 0.9$ TeV & (80 \% \textit{eff}.) & $\pm 0.0365$ & $\pm
0.0054$ & 4.859 & & & & &
\\
$\mid \eta \mid <$ 0.5 &  & & & (23) & & & & &
\\
 & & & & &
\\
\hline
 ALICE & 8550.0 &
1.3764 & 0.2767 & 0.828 & 70.37 & 0.728 & 0.381 & 2.458 & 0.863
\\
$\sqrt{s}= 0.9$ TeV & & $\pm 0.0318$ & $\pm 0.0051$ & 19.88 & & & &
&
\\
$\mid \eta \mid <$ 0.5 & & & & (24) & & & & &
\\
 & & & & &
\\
 ALICE & 8550.0 &
1.4316 & 0.1625 & 2.453 & 5 $\cdot 10^{-5}$ & 1.718 & 0.948 & 11.010
& 2.530
\\
$\sqrt{s}= 0.9$ TeV & & $\pm 0.0272$ & $\pm 0.0029$ & 103.01 & & & &
&
\\
$\mid \eta \mid <$ 1.0 & & & & (42) & & & & &
\\
 & & & & &
\\
 ALICE & 8550.0 &
1.4955 & 0.1332 & 3.458 & 7 $\cdot 10^{-13}$ & 2.213 & 1.276 &
15.201 & 3.664
\\
$\sqrt{s}= 0.9$ TeV & & $\pm 0.0271$ & $\pm 0.0024$ & 165.97 & & & &
&
\\
$\mid \eta \mid <$ 1.3 & & & & (48) & & & & &
\\
 & & & & &
\\
\hline
 UA5 & 8550.0 & 1.7987 & 0.1385 & 1.099 & 28.94 & 0.362 &
na & na & 1.14
\\
$\sqrt{s}= 0.9$ TeV & (80 \% \textit{eff}.) & $\pm 0.0319$ & $\pm
0.0024$ & 57.16 & & & & &
\\
$\mid \eta \mid <$ 1.5 & & & & (52) & & & & &
\\
 & & & & &
\\
 \botrule
\end{tabular} \label{Table3}}
\end{table}

To find out what is the reason for the above-mentioned disagreement
the calculations of Table~\ref{Table2} have been repeated for ALICE
measurements at $\sqrt{s}= 0.9$ TeV but with the arbitrary
assumption that all cases have the same number of events as UA5
ones. The results are listed in Table~\ref{Table3}. One can see that
now there is full agreement between $\chi^2$ and $\chi_{LS}^2$ test
statistic results for all ALICE cases. This means that the accuracy
with which experimental distributions approximate the NBD has not
increased in ALICE data even though the sample sizes are one order
greater. But the accuracy should increase with the sample size
because if the hypothesis is true the postulated form of
distribution is exact for the whole population. So with the growing
number of events, the experimental distribution should be closer to
the postulated one. This is also seen in the form of $\chi_{min}^2$,
Eq.~(\ref{FinalChi}), where the linear dependence on $N$ is
explicit. To keep $\chi_{min}^2$ at least constant when $N$ (the
sample size) is growing the relative differences between $P(Y_i)$
and $P_i^{ex}$ have to decrease.

\section{Conclusions}
\label{Conclus}

The main conclusion is that the hypothesis of the NBD of
charged-particle multiplicities measured by the ALICE Collaboration
in proton-proton collisions at $\sqrt{s}$ = 0.9 and 2.36 TeV should
be rejected for all pseudo-rapidity window classes. This is the
result of likelihood ratio tests performed for the corresponding
data samples. The significant systematic errors are the reasons for
acceptable values of the least squares test statistic for the
narrowest pseudo-rapidity window measurements.

The second conclusion is that the size of ''proper'' errors ({\it
i.e.} not too big and not too small, both extremes cause the false
inference from $\chi_{LS}^2$/$n_d$ values) is somehow related to the
sample size. Here, for instance, errors of the type $\sqrt{n_i}$
could be ''a frame of reference'' as it has been revealed from the
results gathered in Tables~\ref{Table2} and~\ref{Table3}. This is
connected with the meaning of the formulation of a hypothesis. If
the hypothesis is true, it means that the form of a distribution
postulated by this hypothesis is exact for the whole population.
Thus for the very large samples (as in all ALICE cases) the measured
distribution should be very close to that postulated. The performed
analysis has shown that the ALICE experimental errors are much
bigger than the acceptable discrepancies (acceptable for these
sample sizes). Therefore $\chi^2$ and $\chi_{LS}^2$ test statistics
give the opposite answers in the narrowest pseudo-rapidity windows
of the ALICE measurements. For the UA5 sample sizes, which are much
smaller than the ALICE ones, the experimental errors have turned out
to be of the order of acceptable discrepancies, so both test
statistics give the same answer.

\section*{Acknowledgments}

The author thanks Jan Fiete Grosse-Oetringhaus for providing him
with the numbers of entries in the ALICE event classes. This work
was supported in part by the Polish Ministry of Science and Higher
Education under contract No. N N202 231837.

\appendix

\section{}

The sample defined in Sect.~\ref{Likmeth} can be divided into $m$
bins with the different value of measured $N_{ch}$ in each bin. Let
$n_i$ be the number of events in the $i$th bin, {\it i.e.} events
with the same measured value of $N_{ch}$, say $Y_i$. Then the number
of events in the sample equals

\begin{equation}
N = \sum_{i=1}^{m}\;n_i \;. \label{Nevents}
\end{equation}
Dividing by $N$ one can obtain the condition for experimental
probabilities (frequencies) $P_i^{ex}$:

\begin{equation}
1 = \sum_{i=1}^{m}\;\frac{n_i}{N} = \sum_{i=1}^{m}\;P_i^{ex} \;.
\label{Freqexp}
\end{equation}

Now the likelihood function, Eq.~(\ref{Likelfun}), can be rewritten
as

\begin{eqnarray}
&&L(\textbf{X} \mid p, k) = \prod_{j=1}^{N}\; P(X_j; p, k) =
\prod_{i=1}^{m}\; P(Y_i; p, k)^{n_i} \cr \cr & & = L(\textbf{Y} \mid
p, k) = \bigg [ \; \prod_{i=1}^{m}\; P(Y_i; p, k)^{\frac{n_i}{N}}
\bigg ]^{N} \cr \cr & & = \bigg [ \; \prod_{i=1}^{m}\; P(Y_i; p,
k)^{P_i^{ex}} \bigg ]^{N}\;, \label{Likfunbin}
\end{eqnarray}
and the corresponding log-likelihood function reads

\begin{equation}
\ln L(\textbf{Y} \mid p, k) = N \sum_{i=1}^{m}\;P_i^{ex}\; \ln
P(Y_i; p, k) \;. \label{LogLikfubi}
\end{equation}
Since the logarithm of the NBD is given by

\begin{eqnarray}
&&\ln P(n; p, k) \cr \cr & &= \sum_{j=1}^{n} \ln{(k+j-1)} + n
\ln{(1-p)} + k \ln{p} - \ln{(n!)}\;, \cr & & \label{LogNBD}
\end{eqnarray}
the necessary conditions for the existence of the maximum,
Eqs.~(\ref{logLikeqp}), have the following form:

\begin{eqnarray}
&&\frac{\partial}{\partial p}\ln L(\textbf{Y} \mid p, k) \cr \cr & &
= N \sum_{i=1}^{m}\;P_i^{ex}\;\bigg [ -Y_i \frac{1}{1-p} +
\frac{k}{p} \bigg ] \cr \cr & & = N \bigg [ -\frac{1}{1-p}
\sum_{i=1}^{m}\;P_i^{ex} Y_i  + \frac{k}{p} \sum_{i=1}^{m}\;P_i^{ex}
\bigg ] \cr \cr & & = N \bigg [ -\frac{1}{1-p} \langle N_{ch}
\rangle + \frac{k}{p} \bigg ] = 0 \;, \label{DlogLikeqp}
\end{eqnarray}
\begin{eqnarray}
&&\frac{\partial}{\partial k}\ln L(\textbf{Y} \mid p, k) \cr \cr & &
= N \sum_{i=1}^{m}\;P_i^{ex}\;\bigg [ \sum_{j=1}^{Y_i}\;
\frac{1}{k+j-1} + \ln{p} \bigg ] \cr \cr & & = N \bigg [
\sum_{i=1}^{m}\;P_i^{ex}\; \sum_{j=1}^{Y_i}\; \frac{1}{k+j-1} +
\ln{p} \bigg ] = 0 \;, \label{DlogLikeqk}
\end{eqnarray}
where the sum over $j$ is 0 if $Y_i = 0$.

From Eqs.~(\ref{DlogLikeqp}) and (\ref{Parametpk}) one can obtain:

\begin{equation}
\langle N_{ch} \rangle = \frac{k(1-p)}{p} = \bar{n}\;.
\label{Peqalaver}
\end{equation}
Expressing $p$ as a function of $k$ and $\langle N_{ch} \rangle$

\begin{equation}
\frac{1}{p} = \frac{\langle N_{ch} \rangle}{k} + 1 \;,
\label{Oneoverp}
\end{equation}
and substituting it to Eq.~(\ref{DlogLikeqk}) the equation which
determines $\hat{k}$ is obtained:

\begin{eqnarray}
&&\frac{\partial}{\partial k}\ln L(\textbf{Y} \mid p, k) \cr \cr & &
= N \bigg [ \sum_{i=1}^{m}\;P_i^{ex}\; \sum_{j=1}^{Y_i}\;
\frac{1}{k+j-1} - \ln{\bigg (1 + \frac{\langle N_{ch} \rangle}{k}
\bigg )} \bigg ] = 0 \;. \cr & &  \label{DlogLikfk}
\end{eqnarray}
The above equation can be solved numerically. Having obtained
$\hat{k}$ and substituting it into Eq.~(\ref{Oneoverp}) $\hat{p}$ is
derived.

\section{Wilks's theorem}

Let $X$ be a random variable with p.d.f $f(X,\theta)$, which depends
on parameters $\theta = \{ \theta_1,\; \theta_2,...,\theta_d \} \in
\Theta$, where a parameter space $\Theta$ is an open set in
$\textrm{R}^d$. For the set of $N$ independent observations of $X$,
$\textbf{X} = \{ X_1,\;X_2,...,X_N \}$, one can defined the
likelihood function

\begin{equation}
L(\textbf{X} \mid \theta) = \prod_{j=1}^{N}\; f(X_j; \theta) \;.
\label{LikelfunB}
\end{equation}
%
Now consider $H_0$, a $k$-dimensional subset of $\Theta$,  $k < d$.
Then the maximum likelihood ratio can be defined as

\begin{equation}
\lambda = \frac{\max_{\theta \in H_0}{L(\textbf{X} \mid
\theta)}}{\max_{\theta \in \Theta}{L(\textbf{X} \mid \theta)}} \;.
\label{LikeliratioB}
\end{equation}
This is a statistic because it does not depend on parameters
$\theta$ no more, in the numerator and the denominator there are
likelihood function values at the ML estimators of parameters
$\theta$ with respect to sets $H_0$ and $\Theta$, respectively.

The Wilks's theorem says that under certain regularity conditions if
the hypothesis $H_0$ is true ({\it i.e.} it is true that $\theta \in
H_0$), then the distribution of the statistic $-2\ln{\lambda}$
converges to a $\chi^2$ distribution with $d-k$ degrees of freedom
as $N \longrightarrow \infty$ \cite{James:2006zz,Hoel:1971aa}. The
proof can be found in Ref.~\refcite{Dudley:2003ln}. Note that $k =
0$ is possible, so one point in the parameter space (one value of
the parameter) can be tested as well.

\end{document}